\documentclass[preprint,12pt]{elsarticle}

\usepackage{amsmath}
\usepackage{tabularx}
\usepackage{longtable}
\usepackage{subcaption}
\usepackage{url}

\journal{}

\begin{document}

\begin{frontmatter}

\author[inst1]{Anca-Simona Horvath}

\affiliation[inst1]{organization={Research Laboratory for Art and Technology; Aalborg University},
            city={Aalborg},
            country={Denmark}}

\author[inst2]{Alina Elena Voinea}
\author[inst3]{Radu Arieșan}

\affiliation[inst2]{organization={Architecture, Culture, Design, Sustainability; Department of Architecture;
Faculty of Architecture \& Urban Planning; Technical University of Cluj-Napoca},
            city={Cluj-Napoca},
            country={Romania}}

\affiliation[inst3]{organization={Projected Space, Produced Space, Perceived Space; Department of Architecture; Faculty of Architecture \& Urban Planning; Technical University of Cluj-Napoca},
            city={Cluj-Napoca},
            country={Romania}}

\title{Bio-crafting Architecture: Experiences of growing mycelium in minimal surface molds}

\begin{highlights}
\item results from a workshop aimed at designing and building triply periodic minimal surfaces, 3D printing them using wood composites and impregnating them with mycelium are presented, material experiences of workshop participants are analyzed together with audience members opinions on mycelium as a material with which they might interact in various everyday products
\item using minimal surface cores with mycelium filling combines the good mechanical properties of minimal surfaces with the insulating properties of mycelium and mycelium binds to wood-based filament
\item experiences of working living materials are stronger than experiences of working with inert materials, and designers mostly report feelings of biophilia, but some also exhibit biophobia
\item the results from the workshop were displayed in a public-facing exhibition to a general audience; audience members were positive about the impact of bio-technologies on everyday life in the future, but have divergent opinions on how much ethical considerations should influence research directions in bio-technology.

\end{highlights}

\begin{abstract}
This study documents a three-week workshop with architecture students, where we designed and 3D printed various minimal surfaces using wood-based filaments, and used them as molds in which to grow mycelium. We detail the design process and the growth of the mycelium in different shapes, together with participants' experiences of working with a living material. After exhibiting the results of the work in a public-facing exhibition, we conducted interviews with members of the general public about their perceptions on interacting with a material such as mycelium in design. Our findings show that 3D-printed minimal surfaces with wood-based filaments can function as structural cores for mycelium-based composites and mycelium binds to the filament. Participants in the workshop exhibited stronger feelings for living materials compared to non-living ones, displaying both biophilia and, to a lesser extent, biophobia when interacting with the mycelium. Members of the general public discuss pragmatic aspects including mold, fragility, or production costs, and speculate on the future of bio-technology and its impact on everyday life. While all are positive about the impact on bio-technologies on the future, they have diverging opinions on how much ethical considerations should influence research directions.
\end{abstract}

\begin{keyword} mycelium \sep digital fabrication \sep minimal surface \sep architectural design \sep material-based design \sep material experience



\end{keyword}

\end{frontmatter}
\section{Introduction}
\label{intro}

Buildings are responsible for large amounts of energy consumption and greenhouse emissions while construction waste is a serious problem for environmental sustainability \cite{EUConstruction} with one of the priorities of the European Commission being to secure and facilitate the green transition for architecture, engineering and construction in the next decades \cite{EUConstruction, EURenovation}. One direction to achieve this is through research and development of new materials and building techniques that are easy to recycle or reuse. 

Over the last decade, various living materials have been explored as materials for architecture and design. Among these, mycelium, the mass of branched, tubular filaments (hyphae) of fungi, is among the most promising \cite{AlmpaniLekka2021, armstrong2023towards}. Mycelium, as a natural material, is 100\% biodegradable and can be cultivated at low costs. It has been used in the textile industry considered an alternative to leather \cite{KNIEP2024}, in the packaging industry, as an alternative to plastics \cite{Rejeesh2022, Pohan2023}, and also in architecture, in trying to create building materials that biodegrade faster, and that are grown rather than mined and transported around the world \cite{Kirdk2022, Stelzer2021, AlmpaniLekka2021, Abdelhady2023}. Several approaches to incorporating mycelium in the creation for the built environment have explored mixtures of composites with mycelium showing architectural elements \cite{Kontovourkis2020, colmo2022remediating, Colmo2020}, entire pavilions \cite{Lesna2020}, or concepts for larger-scale projects \cite{Lipinska2022}. Mycelium can be a cheap, natural, and biodegradable material with applications for architecture and construction. However, while mycelium exhibits good thermal and acoustic insulation properties, it has low mechanical strength \cite{ATTIAS2020, Butu2020, ghazvinian2022basics}. This has led \cite{Ozdemir2025} to propose wood-based elements as stay-in-place reinforcement for mycelium-based composites. 

Digital fabrication affords the construction of complex geometries, allowing the investigation of complex structures found in nature that exhibit interesting mechanical properties. Examples of such geometries include triply periodic minimal surfaces. These are interesting for architecture because they minimize the material necessary to cover a certain surface, meaning they minimize material use while maintaining structural integrity. While minimal surfaces have been used in architecture in several projects from the works of Felix Candela to the contemporary TheVeryMany, to our knowledge, no study has been reported where minimal surfaces were used as moulds in which to grow mycelium composites - which is what we report on in this paper.

But novel materials don't only need to be understood in terms of their physical properties, they also need to be accepted by the general public as well as by designers who employ and work with them \cite{Ruhse2025, Karana2015}. Materials take decades before they enter production cycles, and aesthetic appreciation of materials dictates whether or not they end up being broadly used \cite{Sauerwein2017, DUARTEPOBLETE2024}.

In continuation of previous research, in this paper we ask the following  questions: \textit{What are the results of growing mycelium in a variety of minimal surface moulds printed using wood filaments?} (RQ1),  \textit{How do designers experience working with mycelium?} (RQ2) and \textit{How do members of a general audience perceive mycelium as a material for designs they might interact with?} (RQ3).

We present a study describing a three-week workshop with 30 architecture students on 3D modeling and 3D printing various minimal surfaces that were then impregnated with mycelium. We describe the results of growing the mycelium in the minimal surface moulds, the experiences of young designers while working with the mycelium as well as opinions from a general audience who interacted with the results of the workshop during a public-facing exhibition as a three-fold contribution. 

In the following section we present related work on minimal surfaces, mycelium for architectural applications and material perception. Next, in Section \ref{MaterialsMethod} we describe the methods used to conduct this study, while in Section \ref{Findings}, we present our findings, discuss them in Section \ref{Discussion}, and provide concluding remarks in Section \ref{conclusion}.

\section{Related Work}
\label{Related}

There are two related areas of research that are relevant to our study: the first has to do with minimal surfaces, their history and their application in architectural design while the second area has to do with mycelium as a novel living material for architecture and design, both from a material properties perspective, and from the perspective of material perception.

\subsection{Minimal surfaces in architecture}

\begin{figure}
    \centering
    \includegraphics[width=0.85\linewidth]{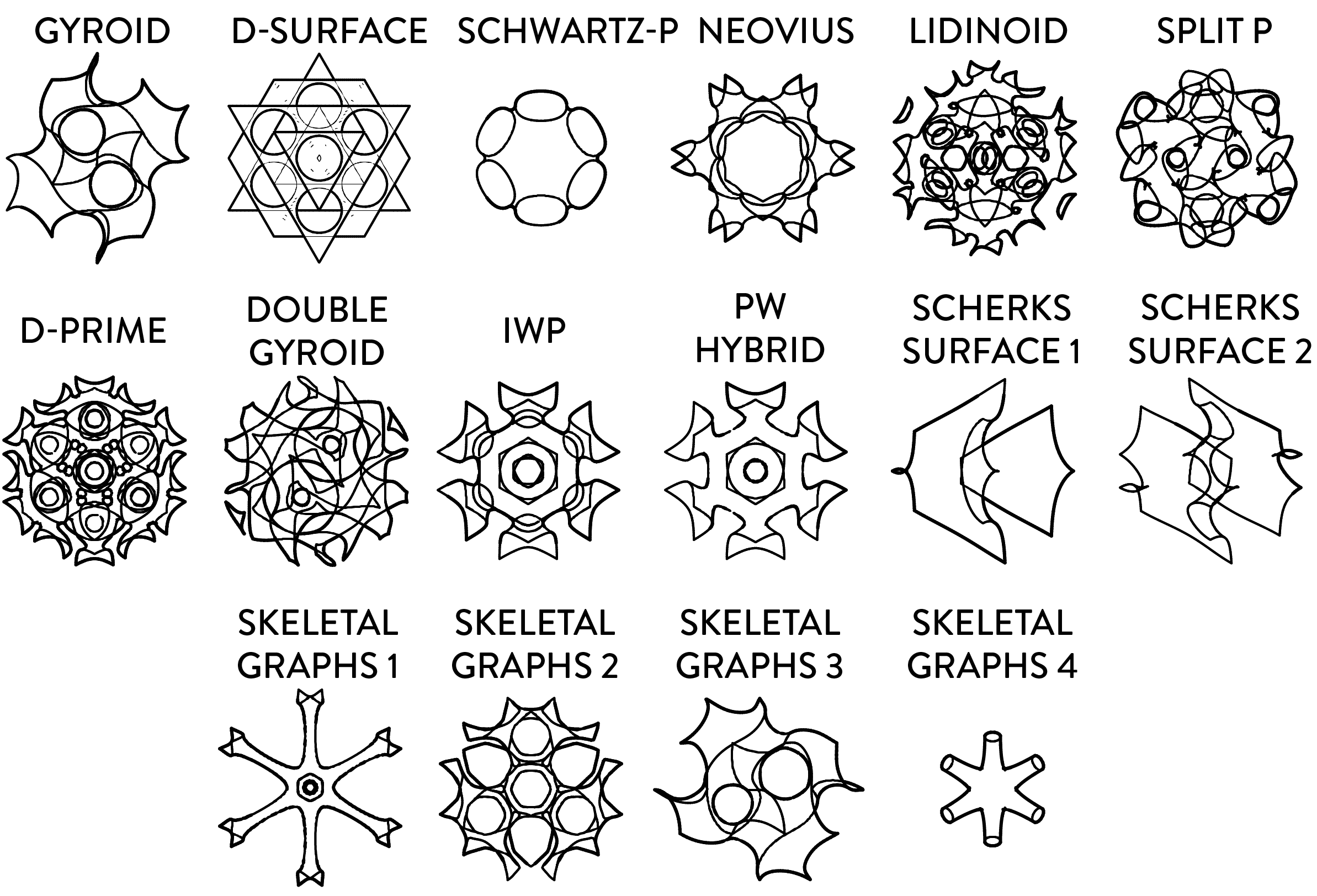}
    \caption{Axonometric views of 16 minimal surfaces.}
    \label{fig:minimal-surf}
\end{figure}

Minimal surfaces represent three-dimensional mathematical surfaces which minimize the surface area for a given boundary with a common example of a minimal surface being soap bubbles \cite{Perez2017}. In mathematics, the first ones to experiment with and propose descriptions of infinite periodic minimal surfaces which do not self-intersect were Schwartz, Riemann and Weisenstrass in the mid 1800s. While there are an infinite number of surfaces that do self-intersect, until 1970, only five triply periodic minimal surfaces had been described \cite{Schoen1970}. In a NASA technical report, Alan Schoen proposed an algorithm that allowed him to describe the five cases already known and twelve additional ones \cite{Schoen1970} - see Fig. \ref{fig:minimal-surf}. Since this discovery, minimal surfaces have been found at micro- and nano- scales in nature, and their properties are very interesting for different applications \cite{Horvath2020}. For example, the gyroid surface was found at nano-scale in the butterfly wings of the \textit{Callophrys rubi} species  where it filters light in such a way that the colors in the wings are given by the way in which the light is filtered, instead of the pigment in the wings. Researchers have found applications for this in photonics \cite{Gan2016}. Due to their unique interaction with light, nano-scale gyroid structures could be used to create compact, light-based electronics where larger numbers of devices can be integrated onto a single chip \cite{Turner2013}.

For architects and designers, minimal surfaces are interesting because they can help build structures using less material and thus contribute to building more sustainably. At the same time, these surfaces have organic and fluid aesthetic qualities which have been used to create iconic architectural designs.
Among the more famous examples of architects interested in and working with minimal surfaces was Frei Otto, an architect and structural engineer who specialized in lightweight tensile and membrane structures. The Munich Olympic stadium, built in 1972, was considered revolutionary for the time and remains an architectural icon today. The canopy of the stadium includes large elements of acrylic glass (of 75x75cm) stabilized using steel cables. These are supported by cable net structures consisting of multiple saddle-shaped surfaces framed by edge cables and suspended from masts. Perimeter masts support primary cables that in turn provide resistance to an interior network of interior cables and flying masts that form the structure of the roof \cite{Schlaich2025}. The project required complex calculations and a computer application was developed specifically for it (an important moment in the history of computer-aided design for architecture). 

Felix Candela is another example of an architect and structural engineer who was interested in minimizing material use and is well-known for developing concrete thin shells. In his work, he often used the hyperbolic paraboloid, also known as a hypar or the saddle surface. While the hyperbolic paraboloid is very close to a minimal surface, from a mathematic point of view, it does not constitute one. Nevertheless, Candela's work is among the most important examples of structural innovation motivated by minimizing material use to cover a certain surface. 

More contemporary examples of using minimal surfaces at large scales and for architectural design include work from the studio Marc Fornes / TheVeryMany who designed and built a number of pavilions, canopies and large scale sculptures. Among them, Minima|Maxima is a permanent pavilion developed for the World Expo in 2017, placed in Astana Kazaksthan and part of a larger series of \textit{Structural Stripes "Crawling Assembly'"} all created from flat aluminium stripes connected with rivets. Minima Maxima takes the shape of an egg and stands at 13m high with an interior of a modified gyroid. This geometry is made out of three layers of flat aluminium strips of 6mm joined by rivets \cite{Fornes2017}. The Orb, commissioned by Google and completed in 2023 is a similar pavilion, only with a spherical exterior and a double gyroid interior structure and built in a similar way to Minima|Maxima \cite{Fornes2023}. The work of Marc Fornes / TheVeryMany makes use of computational design techniques to rationalize these complex structures, and of digital fabrication to precisely cut the stripes but also to assemble the structures.
Another example of the use of a minimal surface for a relatively large-scale project is the Hypar Up pavilion, where wood scraps were used to create a structure of the minimal surface Schwartz-D (also known as Diamond surface) \cite{Colabella2024}. Vlad Țenu has also designed and exhibited a number of sculptural installations of gyroids among them being Nucleotida, Corola or Miniplex \cite{Tenu2025}. These structures were discretised using computational design techniques into flat elements that were later assembled.

Apart from these structures that propose using minimal surfaces at larger scales, there is a growing number of examples where minimal surfaces are used at smaller scales and find applications in design, including aeronautics and aerospace for instance, where creating elements that are as light as possible is paramount. For example, \cite{drones2022} describe using the gyroid (one of the most studied minimal surfaces) as interior structure for airplane wings, where the gyroid helps reduce material use and weight of the wing while maintaining their structural integrity.

Regardless of the scale of the minimal surfaces, these geometries are almost always built making use of advanced computational design techniques: for the larger scale structures, the minimal surfaces are often discretised into flat elements (which are then laser cut). For smaller scale structures, 3D printing is frequently used. Moreover minimal surfaces are often options for interiors (or infills) of 3D prints in 3D printing slicer software (software applications used to prepare 3D models for 3D printing), because of their property of reducing material use while increasing mechanical strength through geometric rather than material distribution. 

\subsection{Mycelium in architecture}

Another strand of research in architecture looks at novel materials for design, and among them, engineered living materials have gained widespread interest across design disciplines \cite{Merritt2020}. Living materials are grown rather than mined (meaning they can regenerate), they biodegrade and in this way can contribute to building more sustainably. Biological materials more generally such as wood, cotton or straw/hay and hemp are examples of materials that have been utilized in building practices for millennia. Out of the newer, engineered-living materials, mycelium is one that has had the highest growth in popularity over the last 15 years \cite{armstrong2023towards}. Mycelium represents a root-like structure that forms the body of a fungus, it can produce mushrooms (which represent the fruits of the fungus) and has the role of collecting nutrients and water. Although fungi have long been associated with plants, they do not photosynthesize and form their own kingdom which is placed between the kingdom of 'animals' and that of 'plants' \cite{Sheldrake2020}. The fungi kingdom is vast, it is considered that up to 90\% of it remains unknown \cite{Sheldrake2020}. Mycelium grows on a variety of substrates including wood, cardboard and even clay. Mycelium can come in the form of a so-called liquid culture containing fungal spores, or in kits made specifically for designers, developed by bio-material companies such as Ecovative \cite{Ecovative2025}, Kineco.bio \cite{Kineco2025} or Grown.bio - in bags with substrates (wood chips, cardboard etc) impregnated with mycelium. These solutions require that they be molded and then grown for 7 to 14 days. Once the fungi has grown into mycelium, it can either be baked - to stop the growth and stabilize the element - or it can be left alive. Leaving the material alive is more efficient in terms of energy use, but might leave the myceliu element more vulnerable to humidity and mould.

Following Phillip Ross' exhibition of Mycotectural Alpha at the Kunsthalle Dusseldorf in 2009 \cite{Ross2009}, architects started to experiment with mycelium in different projects. More broadly across design disciplines, mycelium has been explored as a leather substitute \cite{vandelook2021current}, packaging material \cite{abhijith2018sustainable}, for homeware products \cite{McGaw2022} or as a structural material \cite{Nguyen2022, Ghazvinian2022} - to give only some examples. Several approaches to incorporating mycelium in the creation for the built environment have ended up in a series of pavilions \cite{colmo2022remediating, goidea2020pulp, Ghazvinian2022, Tan2022}, impregnating trimmed wood with mycelium placed in moulds \cite{ozdemir2022wood}, or impregnating 3D printed wood structures \cite{alima2022bio}. In \cite{alima2022bio}, the authors describe a setup where a robotic arm is programmed to inject a mycelium liquid culture into various geometries that were 3D printed using wood filaments. The study finds that the mycelium grows most rapidly along smooth porous surfaces that provide a series of micro-valleys for the organism to seep through \cite{alima2022bio} - this means that other complex geometries as bases on which to grow mycelium should be explored.

Research on mycelium for architecture conducted so far shows that while mycelium is a good candidate to replace polysterene and has good thermal and acoustic properties, it has low mechanical strength \cite{Ghazvinian2019A, elsacker2019mycelium} and so combing complex geometries which have good mechanical strength as bases on which to grow mycelium appears promising - which is what we explore in this study.

The use of mycelium together with digital fabrication technologies to produce elements for buildings is a small but growing field. The hope is that mycelium-grown composites will be more sustainable than the building materials in widespread use today, such as concrete, steel or polystyrene. 

Nevertheless, there are still unknowns when it comes to understanding how new biological materials are perceived by audiences and challenges for mycelium use in the building industry are \textit{psychological, aesthetic} and \textit{economic} rather than technical \cite{McGaw2022, Lipinska2022}. This makes it important to also focus on how mycelium as a material is perceived and on its aesthetic qualities. According to art historian Monika Wagner, materials carry socio-cultural meanings of their own, and not under the control of the artist making use of them \cite{Wagner2002}, they embody ideals and beliefs and drive us to behave in certain ways towards them \cite{Giaccardi2015, Karana215a}. Karana has theorized the concept of material experience \cite{Karana215a}, proposing that, they suggest elicits three kinds of experience during user-material-product interaction: gratification of senses, conveyance of meanings, and elicitation of emotions, and that materials experience should be formalized in design educations \cite{Pedgley2016}. 

To summarize, our research is motivated by the following: (1) minimal surfaces are interesting for architectural design as they minimize material use while maintaining good mechanical strength and they can be used to reduce material use through geometric distribution. (2) Mycelium as a material for architectural design has been growing in popularity and it appears that mycelium growth is influenced by the geometry it is grown on, combining the good mechanical properties and the complex geometries of minimal surfaces with the good insulating properties of mycelium can be a promising research direction. (3) Understanding how audiences perceive living materials such as mycelium and how designers find working with living materials is an area that that requires further investigation as it has been shown that how people perceive a material will influence whether a material will be widely used.

\section{Materials and Method}
\label{MaterialsMethod}

In continuation of previous work, and bridging the three areas identified above, we conducted a workshop, with students of architecture, where we designed and 3D printed various minimal surfaces, using wood-based filaments. Next, these surfaces were impregnated with mycelium, and allowing the mycelium to grow for one week - they were baked to stabilize the mycelium. The results of this process were exhibited in a public-facing exhibition, to a general audience in January of 2024. Participants in the workshop completed a questionnaire on their experiences of working with mycelium, and three semi-structured interviews with members of the general public were conducted. We describe each of these steps in detail in this section.

\begin{figure}
    \centering
    \includegraphics[width=0.9\linewidth]{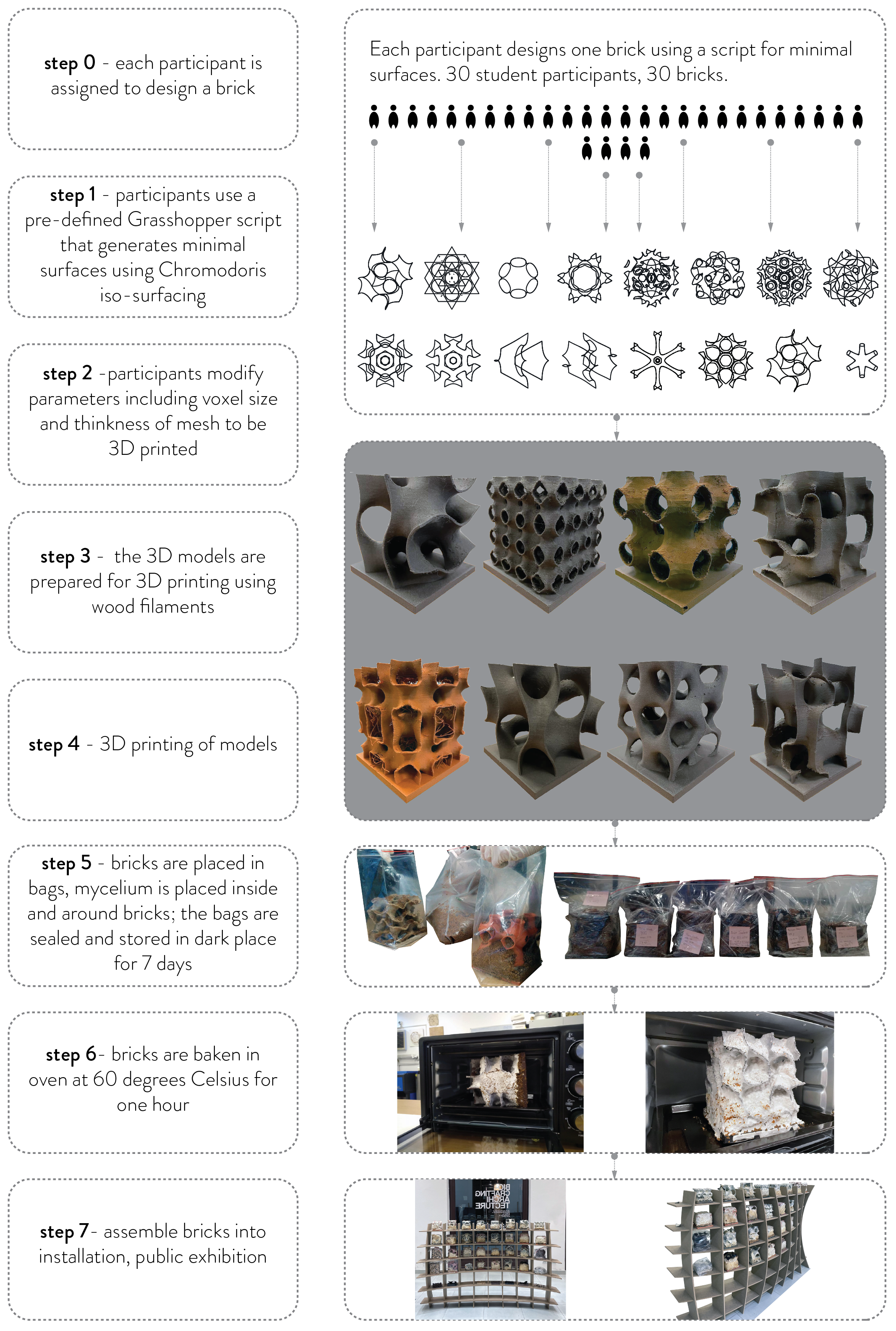}
    \caption{The steps taken during the workshop}
    \label{fig:process}
\end{figure}

The workshop took place over three weeks and 30 students enrolled in an \textit{Architecture and Urban Planning} program in their 3rd and 4th years of study took part. Fig. \ref{fig:process} shows the entire process in detail. The first week was dedicated to creating the 3D models of the minimal surfaces and preparing them for fabrication using 3D printing (steps 0 - 3 in Fig. \ref{fig:process}). The second week was dedicated to 3D printing the bricks (step 4, in Fig. \ref{fig:process}) - once a brick was ready, it was placed in a bag, and impregnated with mycelium (step 5, in Fig. \ref{fig:process}). The last week was used to allow the mycelium growth, and after each brick had spent 7 days growing, it was baked (step 6, Fig. \ref{fig:process}). Each of these steps is described in detail below.  

\subsection{Creating the Minimal Surfaces}
The first step was to create 3D models of minimal surfaces (step 1 in Fig. \ref{fig:process}). This was done using the Grasshopper environment, a visual programming language on top of 3D modeling software Rhinoceoros 3D. After an introduction to general 3D modeling and digital fabrication, as well as to Grasshopper, students were given a script that creates minimal surfaces using the iso-surfacing algorithm implemented in the Chromodoris add-in \cite{Chromodoris2025}. Students could choose the minimal surface they wanted to work with from: gyroid, diamond (or D-surface), Schwartz-P, Neovius, Lidinoid, Split P, D-Prime, double gyroid, IWP, PW hybrid, Scherks surface 1, Schercks surface 2 or skeletal graphs 1-4 (see Fig. \ref{fig:minimal-surf} showing axonometric views of each of these surfaces). The script allowed modifying the surface parameters based on their mathematical functions. The 3D model design had constraints related to their fabrication - meaning they should be 3D printable using the machines we had at our disposal. Each model was designed to fit in a 15x15x20cm volume with the base at 15x15cm and with a 1cm base that would allow better printer bed adhesion. 

\subsection{3D Printing of Minimal Surfaces}

\begin{figure}
    \centering
    \includegraphics[width=0.75\linewidth]{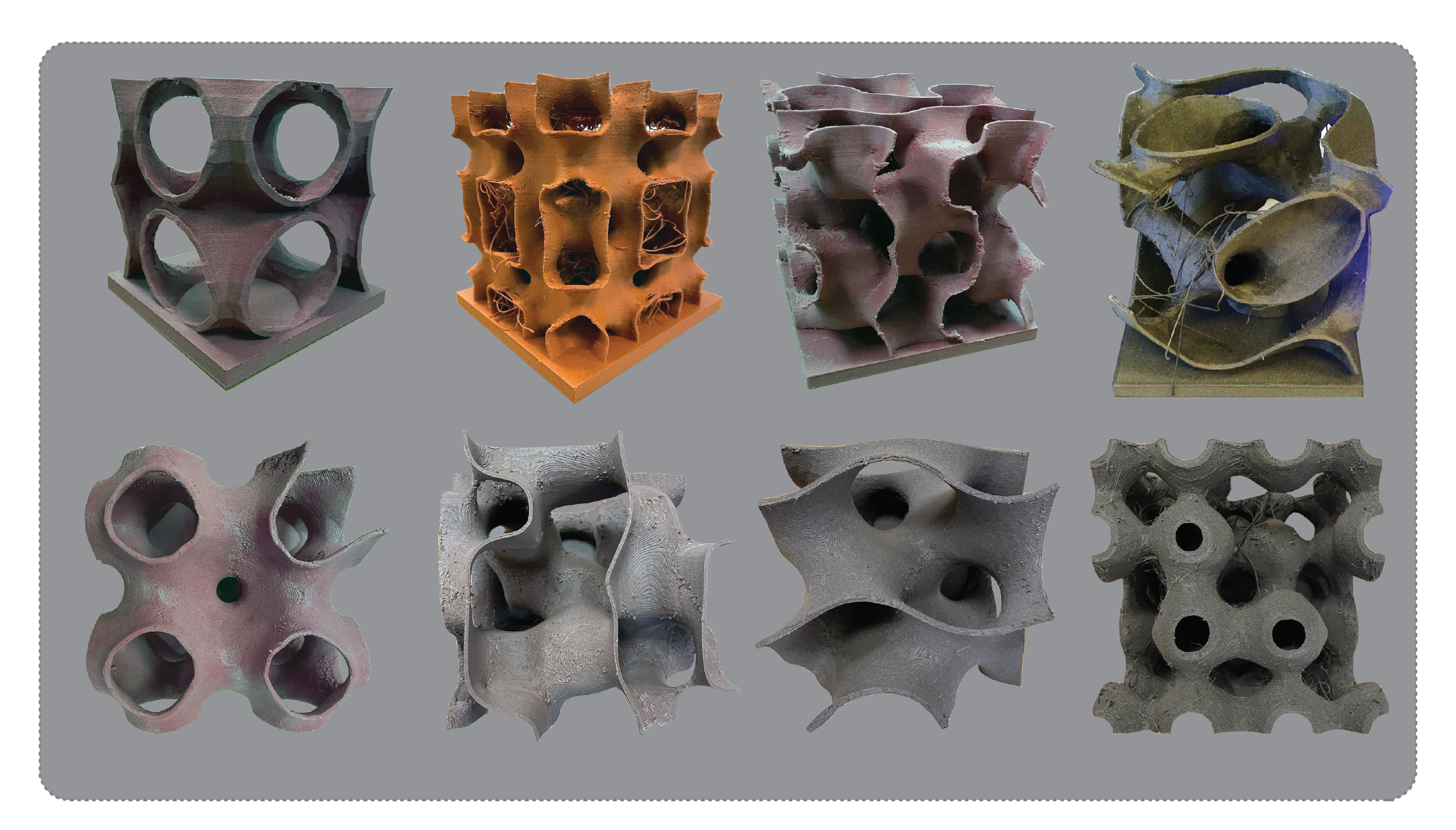}
    \caption{3D printed minimal surfaces before being impregnated with mycelium.}
    \label{fig:printed-bricks-1}
\end{figure}

Once designed, the minimal surfaces were 3D printed using fused deposition modeling using Creality Ender 3D printers and with wood filaments from Timberfill \cite{Timberfill2025} (see Fig. \ref{fig:printed-bricks-1}). These filaments are made out of different types of plastics mixed with natural fibers obtained from wood (accounting for 20\% of the total material composition). According to the material specification sheet, Timberfill is 100\% bio-based and biodegradable in anaerobic conditions and in water \cite{Timberfill2025}. We used 0.6mm nozzles for printing the bricks as we found that smaller nozzles clogged the machines often.

\subsection{Mycelium Impregnation, Growth, and Bake}

We used mycelium from the company Kineco.bio \cite{Kineco2025} that came in 10kg bags of pre-impregnated residual sawdust called hedelcomposite.
After the 3D prints were complete, we disinfected the bricks and placed them in bags with zip ties.  Mycelium was placed in and around the bricks together with water. Students used gloves when working with the mycelium to protect it from infection. After closing the bags and poking some air holes into them, the bricks were placed in a dark room at 22-26 degrees Celsius. After seven days, the impregnated bricks were taken out of the bags and baked in an electric oven, at 80 degrees Celsius, and for 60 minutes. All of these steps follow the instructions from Kineko. In fig. \ref{fig:process} steps 5 and 6 show images from this stage.

\subsection{Public Exhibition}

\begin{figure} [!h]
    \centering
    \includegraphics[width=0.95\linewidth]{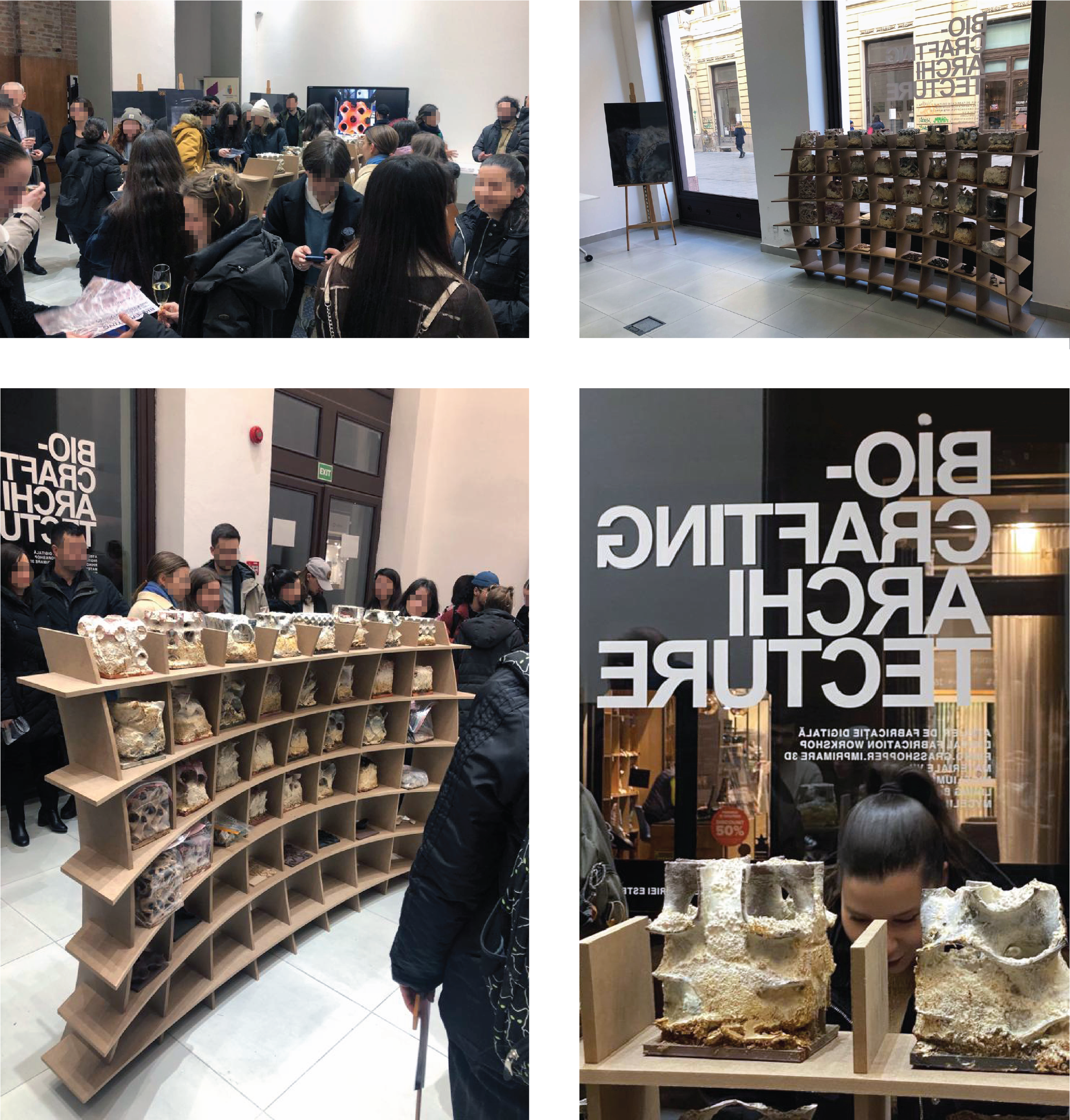}
    \caption{Exhibition of workshop results including images showing the opening vernisage.}
    \label{fig:exhibition-1}
\end{figure}

\begin{figure}[!h]
    \centering
    \includegraphics[width=0.8\linewidth]{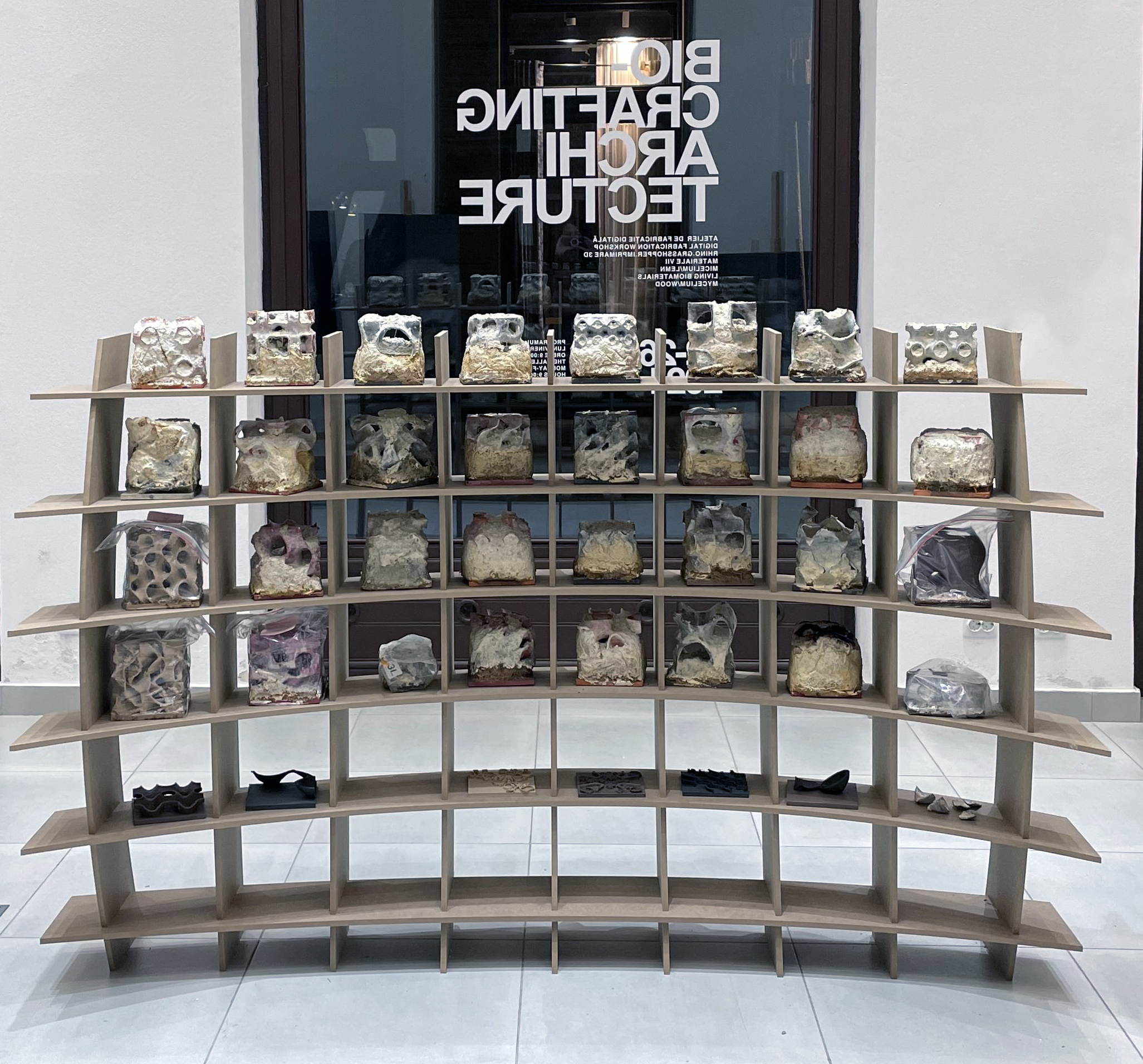}
    \caption{Exhibition of workshop results.}
    \label{fig:exhibition-2}
\end{figure}

Following this, we organized a public exhibition in a cultural venue with the support of the local municipality. This exhibition lasted for 14 days and the public had free access to it. Apart from displaying the mycelium bricks, close-up photographs of some of the pieces were shown as well as a video describing our process. The opening of the exhibition gathered around 60 guests, and during the following days, roughly 100 more people came to see the exhibition (see Fig. \ref{fig:exhibition-1} and Fig. \ref{fig:exhibition-2}).

\subsection{Participant and audience data collection and analysis methods}

After this entire process was completed, we distributed a survey to the participants in the workshop to gain insights into their experiences of working with a living material. The survey was anonymous and consisted of open-ended questions related to their perceptions about mycelium, the opportunities, and challenges they see related to it also as it compares to other materials they might have experience with, how they see the future of architecture and design in general, and as it connects to the development and introduction of new materials. All 30 participants in the workshop completed the survey.

In addition to this, one of the authors conducted semi-structured interviews with three audience members discussing their understanding of and feelings about interacting with living materials for designed products. The interviews lasted between 30 and 45 minutes, were conducted online, and were transcribed by the same author. We printed flyers that were placed in the exhibition venue where we invited audience members to participate in an interview focusing on their perception of bio-materials for design and their opinions on bio-technologies in general. The three audience members visited the exhibition and volunteered to take part in the interview by contacting one of us. 

Both the workshop participants and the audience members signed consent forms as issued by our university, and were informed that their data would be anonymized, stored in a GDPR compliant way, and deleted after 2 years. Moreover, they could opt-out from participating in the study at any time. None of them chose to do so. 

Both the data collected from workshop participants and that collected from audience members were analyzed qualitatively in two separate processes, using a thematic analysis approach \cite{Naeem2023}. In each of these cases, two of the authors printed the data and spent time individually becoming familiar with them and later coded them by using an emergent coding approach \cite{Lazar2010}. The codes were then negotiated between the two authors until we agreed on final lists for each of the datasets. Afterward, an iterative process started where the codes were affinity diagrammed \cite{Holtzblatt}, first separately and then collaboratively by two authors until final structures were produced for each dataset. Based on these two analyses, responses on perceptions of living materials and design were distilled into main themes used to structure the presentation of the data in the following section.

\section{Findings}
\label{Findings}

\begin{figure}[!h]
    \centering
    \includegraphics[width=0.85\linewidth]{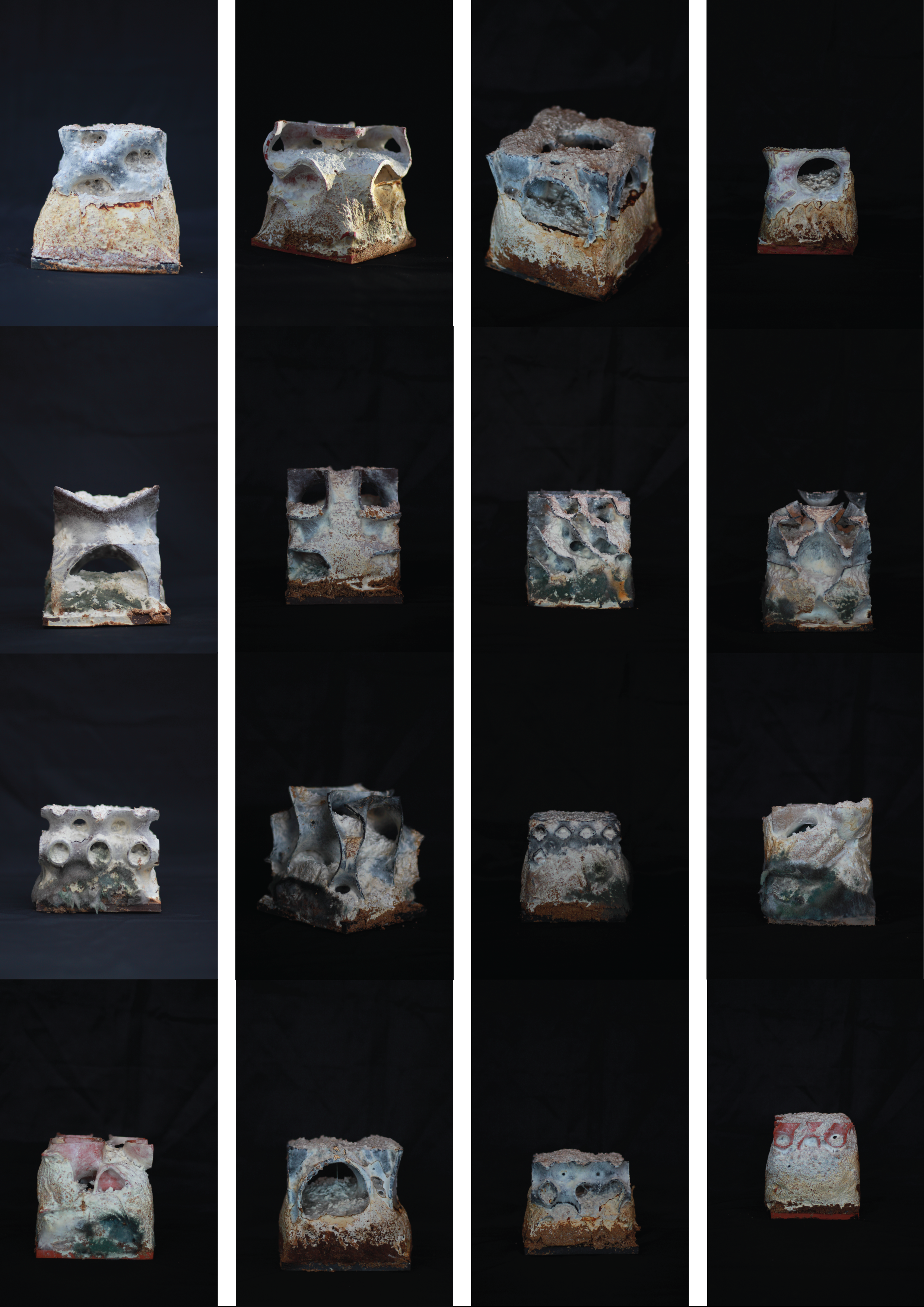}
    \caption{The resulting bricks after being impregnated with mycelium and baked, and after two weeks in storage.}
    \label{fig:individual-bricks-1}
\end{figure}

\begin{figure}[!h]
    \centering
    \includegraphics[width=0.95\linewidth]{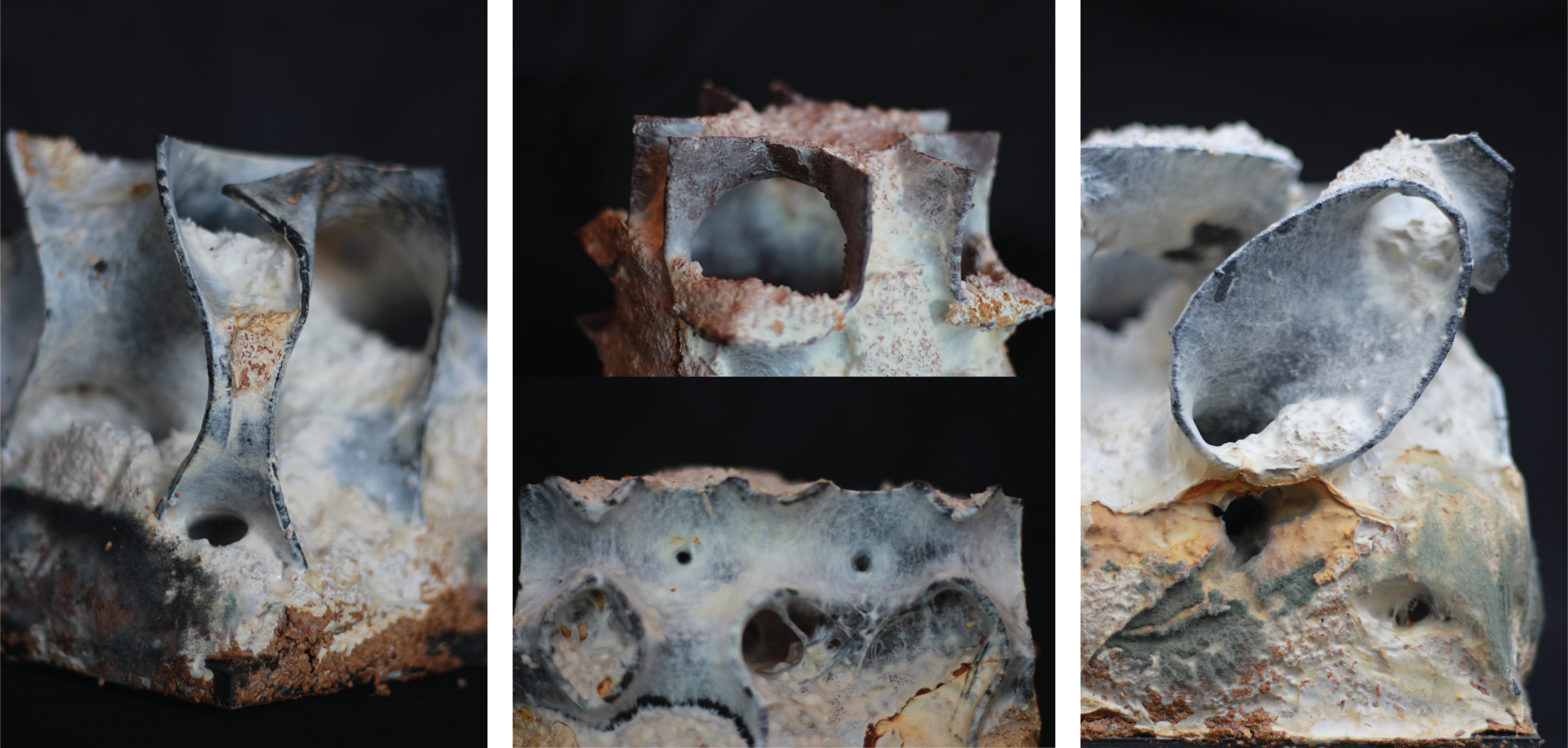}
    \caption{Details of resulting bricks after being impregnated with mycelium, and baked.}
    \label{fig:individual-bricks-3}
\end{figure}

\begin{figure}[!h]
    \centering
    \includegraphics[width=0.95\linewidth]{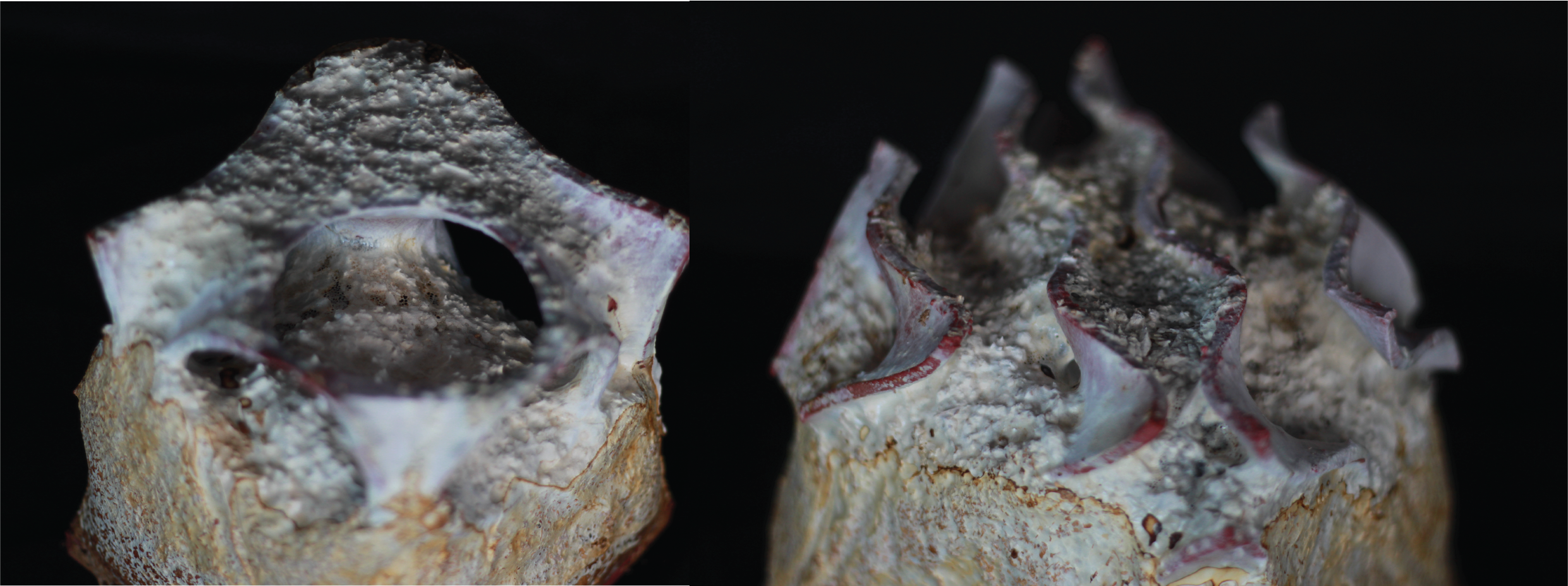}
    \caption{Details of resulting bricks after being impregnated with mycelium, and baked.}
    \label{fig:individual-bricks-4}
\end{figure}

\begin{figure}[!h]
    \centering
    \includegraphics[width=0.95\linewidth]{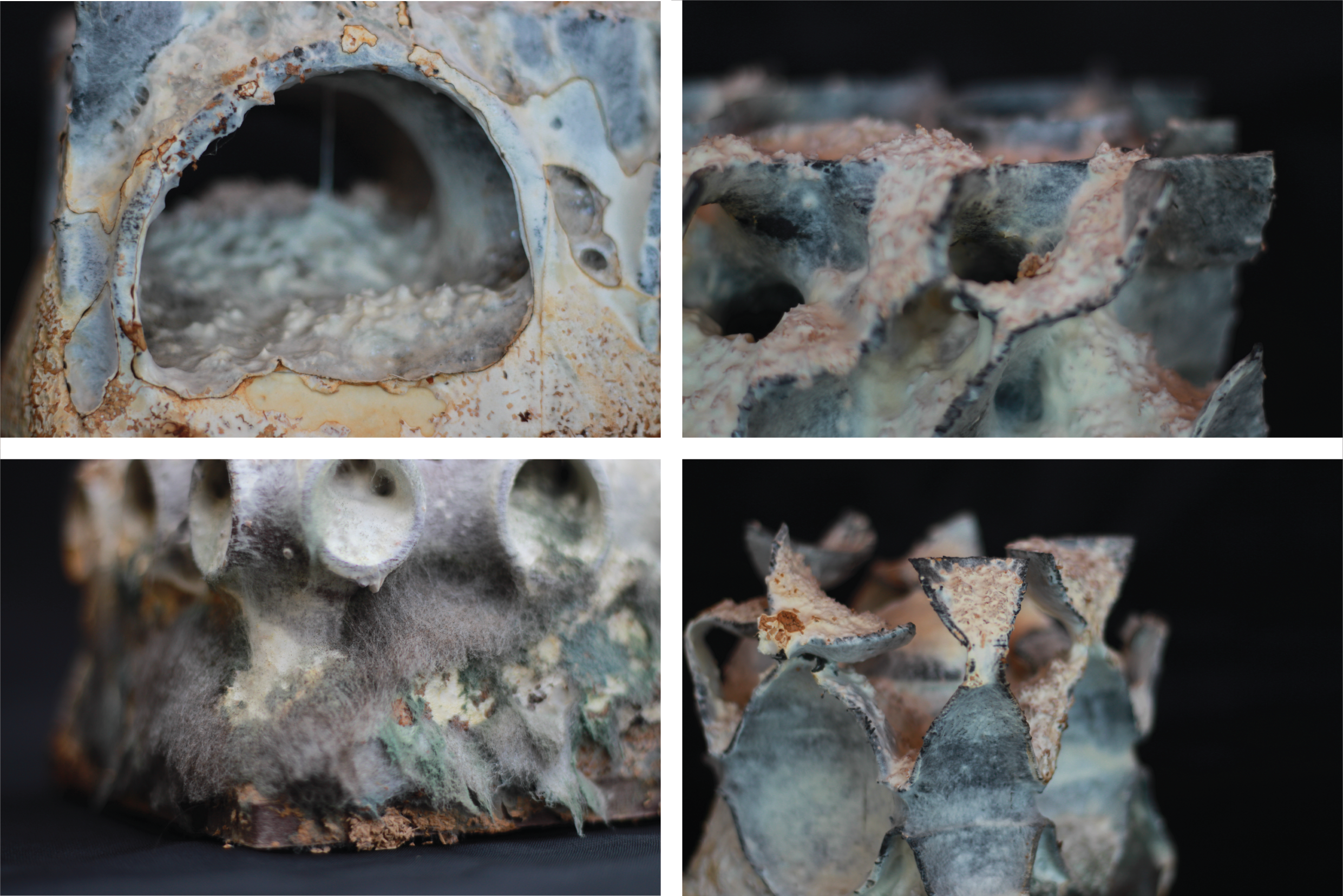}
    \caption{Details of resulting bricks after being impregnated with mycelium, and baked.}
    \label{fig:individual=bricks-5}
\end{figure}

\begin{figure}[!h]
    \centering
    \includegraphics[width=0.8\linewidth]{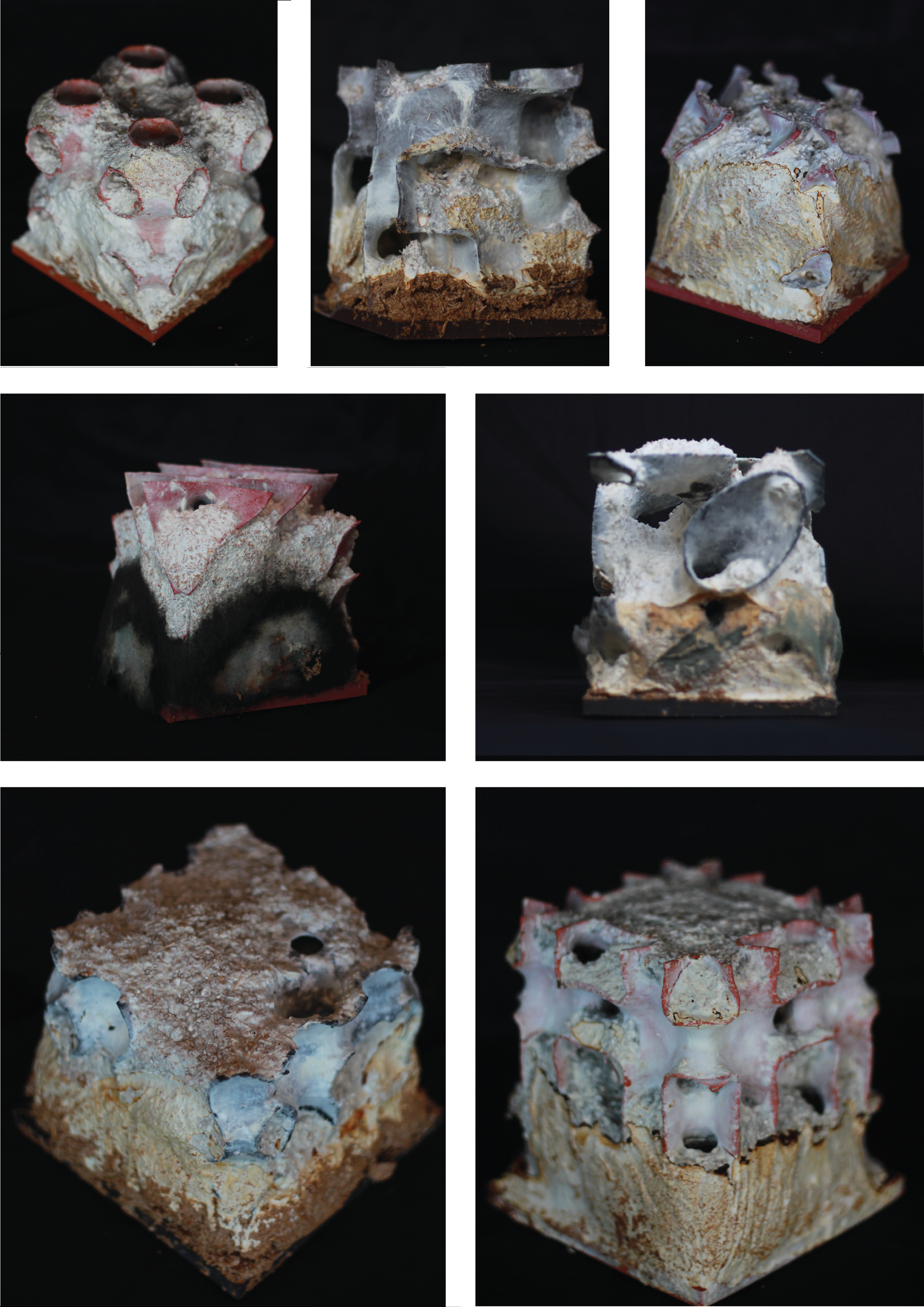}
    \caption{The resulting bricks after being impregnated with mycelium and baked.}
    \label{fig:individual-bricks-2}
\end{figure}

Our findings are presented below by following the three research questions we set out to answer: first, we show the results from impregnating the 3D-printed minimal surfaces with mycelium. Next, we present designers' experiences of working with mycelium in their design processes. And finally, we introduce the results from the semi-structured interviews with audience members from the general public.

\subsection{Mycelium impregnated bricks}

By conducting visual and photographic analysis of the resulting bricks, we notice that combining minimal surface cores with mycelium filling, we see the mycelium grew on the bricks and stuck to them as the hedecopter mycelium blended with the bioplastic-wood composite from the Timberfill filament and the combination resulted in compact elements. 
This means that the mechanical properties of minimal surfaces could be combined with the good thermal and acoustic insulation behavior of mycelium to create composite elements, and research to scale up this approach can be promising.

Mold became a problem in almost all of the bricks. Mold set in on the bricks after spending time in a dark storage room and before being exhibited to the general public. This can be mitigated by: (a) baking the bricks for a longer time and thus reducing the water content of the mycelium (although this will increase the total energy to be used for producing such elements); (b) controlling the environmental humidity in which the bricks are employed, although this might not be feasible. 

\subsection{Designers' material experiences of working with mycelium}

All participants were aged between 21 and 23, apart from one who was 27 at the time of the workshop, the participants are numbered P1 through P30 for the quotes below. All of them had experience with at least one Building Information Modeling software, the most common being ArchiCAD and Allpan Nemetschek, while some also mention Revit. Apart from this, Sketchup is a popular software along with 3DS Max, and Lumion, and a handful reported using Blender, Enscape, Twinmotion, and Nomad (for sculpting). Only one of the participants had done a conceptual project that proposed the use of mycelium for an interior design project in the past, but they did not have the chance to work hands-on with the material. The rest had no experience with living materials before the workshop.

Describing their experience of working with the mycelium, 14 of the participants used the word \textit{interesting} in their comments, while others used words such as \textit{fantastic} (P8), \textit{unusual} (P15), \textit{novel} (P23), \textit{new and exciting} (P13) or even \textit{eye-opening} (P23) and \textit{revolutionary} (P29). Many state that they \textit{like} the mycelium, and that they find it to be \textit{friendly} (P6, P7, P10, P17, P18, P21). According to P10: \textit{I liked to see how it changes from one day to another and how it takes the shape of the 3D model I created}. P14 states that: \textit{It was really exciting: I was so amazed by the way our pieces turned out} while, while for P9: \textit{It was really exciting: I had no idea how to imagine a living material and I was a little bit afraid [...], but afterwards I was sure I was working with something that has huge potential for the future}. For P18, \textit{it was interesting: both the process and the result were unexpected; I really enjoyed to work with mycelium, it did not feel difficult to work with, it was exciting to see its transformation}. Similarly, P15 illustrates their experience as \textit{unusual, I was excited about seeing the final result} and for P26: \textit{the process by which the mycelium transformed over time was very interesting as well as the way in which it took over our 3D model} and similarly, in P13's case: \textit{At the beginning of the process I wasn't sure about the result, and how it will turn out, so it was a surprise seeing the final product [...] The fact that it changed its dimension, texture and colour was very exciting, not being sure how it will look at the end}. But while some are intrigued and excited about the qualities of the mycelium as a living, unpredictable material, others are slightly less enthusiastic about not knowing what to expect from it: \textit{It was a great idea, but in the end, the shape that I created cannot be seen because of the mycelium. Now, mine is a full cubic block.} (P1)

On the other had, few participants describe feeling more ambivalently (a total of five from the 30 participants), and words such as \textit{scary}, or descriptions of feeling \textit{afraid} were used: \textit{the experience was interesting, but it was also a bit scary knowing I had to disinfect my hands afterwards and then watch it overtake the model I made} (P3). According to P24: \textit{Working with mycelium was a very interesting experience: on one hand because of the need to protect ourselves from it, on the other because of lack of control upon its exact final shape, both unlike any other materials I have previously worked with.} P19 was the only participant to portray a more negative encounter: \textit{It was an interesting experience, however the smell was awful and it caused me itchy skin.}

\subsubsection{Mycelium compared to non-living materials}
Mycelium is often compared to other materials they had used before, and here most participants discuss \textit{unpredictability}.
For some this is acceptable and even exciting, or simply something to accept as a property of the material, while for others it is irritating, or frustrating. P10 writes: \textit{If we want to create a regular 3D printed item, we just buy the 3D printing filament, design and print the object and that is it. But if we want to use mycelium, we have to [...] wait for it to grow, bake it - it takes time to get to the final product}.  Similarly P4 explains that: \textit{you don't have any control on the shape it chooses to take} while P3 tells that: \textit{you cannot control the process as easy, you let it make some decisions in appearance, design and the final product is a teamwork of sorts.} P11 is bothered by this lack of control: \textit{preparing everything involves much more work and sometimes it can be irritating when you cannot fully control the growth of the mycelium. Sometimes during the process you have to be very precise - however, the great advantage is that it is a natural material, that grows fast and is environmentally sustainable.}

Nevertheless, some find the livingness particularly appealing: \textit{it was pretty impressive to see your sculpture grow} (P14) while P29 says that: \textit{It's like bringing life!}.  A few mention the \textit{responsibility} of the designer to \textit{care} for the material is something they enjoy. For example, P19 describes: \textit{You are responsible of it, just like a plant, it has to be taken good care of}, while P20 writes: \textit{it is a growing being, you have to care for it!}  Similarly, P21 explains \textit{you don’t want it contaminated, so you need to be extra-careful, you need to disinfect everything before and after, something that with other materials would be not necessary}.

\subsubsection{Ethics and using using living materials for design}
Ethics was also among the themes discussed, and here the opinions are almost equally divided between those that see no ethical issues involved in designing with mycelium - as they do not consider it a sentient being, and it is similar to foods we eat, and those that think that ethics should be considered as looking more holistically at the environment and ways in which human intervention can modify ecologies. For example, P3 believes that \textit{I think ethics ends at sentience}, while P21's states: \textit{I see no ethical issues as long as we don’t involve any sentient creature in our designs}. P17 puts it more bluntly: \textit{I don’t see any negative implications as long as the material does it’s job responding to construction standards}. However, others believe that designers should consider ethics when engaging with living materials, and that it is especially important to \textit{respect nature}. For P10: \textit{designing with living materials raises ethical concerns, including issues related to the treatment of living organisms, their environmental impact, and the responsibility to prioritize ecological balance}; while P16 says that: \textit{we need a deep respect for the inherent value of life.} Similarly, P24 thinks that \textit{using living materials in architecture raises ethical considerations regarding the well-being of the ecosystem}. All these comments are similar to P30's opinion: \textit{ethical considerations need to ensure we don't cause any harm to the living materials and to the ecosystems they live in}. In short, P22 puts it: \textit{we have the ethical responsibility to set limits in ways in which we modify nature}. 

\subsubsection{Looking into the future of architecture and design}
Many reflect on design futures and the responsibility of design professionals within this context where living materials are generally seen as part of the future of design. Along with this, advanced technologies, and especially AI's influence on the profession was often mentioned. This is perceived with mixed attitudes, between believing that AI will help make work more efficient and fast, but also with a sense of anxiety over its impact on workers: \textit{I wonder what will happen if an AI could make your dream house in minutes. But AI doesn't have what we have: soul and feelings. It can apply rules, but if you don't design with your heart also, then there's no difference between you and a programmed robot} (P1). Similarly P5 writes that \textit{no machine can replace the brain of a creative person such as an architect} while P9 hopes that \textit{not everything will be done by AI}. The future is almost always discussed alongside sustainability concerns. Some discuss re-use of existing buildings, and fewer demolitions, together with research on new materials: \textit{we are going to have a huge responsibility in reusing existent buildings and integrating sustainable materials into the whole industry} (P9) and \textit{we know architecture has a big environmental footprint and there are already a lot of solutions that could be applied. We will build fewer buildings from concrete and glass, these materials will be replaced with more sustainable ones. Architecture will be more about society and spaces will be designed and thought through with much more care} (P11). For P12: \textit{architecture will be based more on the reconstruction and rehabilitation of existing buildings than designing and building something new}. For P22: \textit{the future means finding replacements to materials that are polluting and hard to recycle}. Finally, according to P4, changing ideologies as well as population growth will need to be considered: \textit{the future of architecture is constantly changing because of the new challenges that come with the growth of population, as well as changes of the ideologies and beliefs}.

\subsection{General audience perceptions of living materials for design}
After exhibiting the results of our work to a general audience, we conducted semi-structured interviews with three volunteers (we call them observers: O1, O2 and O3) aged 35, 37 and 67 respectively. O1 is a PhD student in genetics engineering, O2 is a theater director and university lecturer also involved in the committee for \textit{Ethics and Academic Integrity} at her university, while O3 is a structural engineer with a lifelong experience in construction projects at different scales. The three discussions were colored by the backgrounds of the participants and took very different directions, however three themes came up in all conversations: \textit{challenges} of designing with life, the \textit{ethics} of designing with life,  and the \textit{future}.

 However, all three declared themselves optimistic and enthusiastic about bio-technologies and their possible applications for design and healthcare in the future. 

\subsubsection{Ethics of working with living materials for design}

O1 believes there should be no limits on such research, and that ethics considerations are limiting research within the field of genetic engineering, thus harming progress. On the other hand, O2 and O3 believe some limits (or regulations motivated by ethical and practical concerns) regarding the ways in which we manipulate sentient life should be considered.

O2 discussed ethics extensively starting from its meaning: \textit{ethics is a question of definition and option - what a project or a group of people understand as ethical might not be universally accepted as being ethical}. She also touches on the concepts of \textit{ideal} versus \textit{pragmatic} ethics explaining: \textit{I would prefer to only eat lab-grown meat because I would prefer that animals, that have not given their consent, are not sacrificed so that I have good protein consumption. But my ideal ethics cannot be reconcile with my applied ethics, because of biological and health reasons.} She also conceptualizes ethics as a filter that people might 'wear' or not, meaning \textit{being able to differentiate between ideal and pragmatic ethics: some things are filtered by the immediate reality and the fact that you are part of a system where people have different values} but also that \textit{the tendency is that we as humans understand ourselves as the superior species, and we conceptualise ourselves as: it is ethical to manipulate life forms that are non-sentient, and sometimes even sentient life}. She believes that ethical questions regarding designing with life should be asked by designers: \textit{the person who initiates this type of relationship in this case of mining, or manipulation - has the responsibility to ask this question, and to think it through [...] that mycelium was sitting there, it didn’t want anything from anyone.} 

O3 sees no ethical considerations to be taken into account regarding the development of new materials for design, apart from that they \textit{need to be healthy for humans}. In his view, limitations on genetically modified organisms should consider ethics only when it has to do with human life (for example cloning of humans).
 
\subsubsection{Perceived challenges around living materials for design}

All three audience members discuss various challenges they perceive around using living materials for design.

O1 had strong feelings about de-regulation of genetically modified organisms in Europe. According to her, genetic modifications have a bad public image because of poor public communication on the topic, and there is public resistance to de-regulation which harms research in the field, and impacts the innovation capacity of Europe more broadly (as the US and Asia have fewer regulations on the use of genetically modified organisms, especially in agriculture). She believes that on the one hand these regulations, and on the other hand this public perception, might affect the introduction of living materials for design and research. 

For O2, the main challenge remains around the ethics of designing with life, where, as mentioned, she believes it is important that designers who initiate the relationships with living materials  think through the ethics surrounding such work. In her words: \textit{sometimes I ask myself if ethics (for this particular case) is not a question that is too sofisticated, or far-removed -  of course its very interesting on an intellectual level - and whether there isn’t a disconnect between the type of thinking that asks this question and a reality in which this question doesn’t make sense because it sits in another existential and conceptual plane. This is the main challenge: to connect these different approaches - so that such a question (about the ethcis of using life as a material for design) would make sense to someone who thinks: the mushroom is there, I eat the mushroom, I eat the sheep and I exploit the land because this is the order of things}.

O3's focused on very practical issues about the application of mycelium or other living materials as materials for construction. He brings forward the challenges of scaling up, the costs of a material, including the costs for sourcing, growing it, and stabilizing it: \textit{if growing takes time, then that is also a cost}. He touched on how people's expectations about construction revolve around durability, but also on the regulations in construction: \textit{you have to be realistic - buildings need to withstand earthquakes, severe weather conditions [...] while these materials are interesting, at the end of the day, a large part of construction works with durable, and strong materials}. He also notes that \textit{in construction we work with materials at scale: they need to be cheap enough, and we need to be able to produce them at scale}. But while O3 sees cost, and fragility as issues for mycelium, meaning \textit{maybe it ends up being feasible only for interior, smaller scale or temporary construction, otherwise, as cores in walls, protected, and that's ok - but mold can be a huge issue}, he believes also that research for new materials is crucial, and that minimal surfaces can hold very interesting applications. According to O3, there can be conflicts between the most environmentally friendly solution and the cheapest one: \textit{maybe in richer countries the most environmentally friendly solution could win in a solution competition, but the developing world - people might have to go for the cheapest solution. This is why new material research needs to also look at the ultimate prices}.

\subsubsection{Bio-technological futures}
All three audience members are enthusiastic about the future of bio-technologies and their potential for design.
O1 gave technical suggestions on other possible biological materials designers might be interested in exploring: \textit{any dry tissue from a plant can become a building material}, continuing to suggest succulents as: \textit{they have a layer that is similar to wax, so that wax could be interesting}. She also believes it would be interesting looking at various waste products from agriculture and exploiting possible uses for them in construction: \textit{something that is used in agriculture a lot and comes out as waste is peat [...] maybe it could be a substrate for mycelium [...] or something that’s called ‘mraniță’ in Romanian - its like manure, but its very very dry, it doesn’t even smell anymore and that is used as a fertilizer}. She also suggests trying to find uses to mineral wool that is a by-product of growing vegetables in greenhouses, which once infused with different fertilizers ends up in landfills and is highly polluting. Finally, she explains that genetic engineering could be used on the mycelium to control when it stops growing and suggests (but states that this needs to be checked) that mold could be resolved for mycelium products by genetically modifying the strand of mycelium used: \textit{mould is very interesting because in some sense its the best-adapted mushroom, you get mold on anything if you have enough moisture. So its clear there are mould spores on anything, and I am convinced that as mold is a very smart mushroom, there should be another mushroom that can eat mold. You could make a solution, that might contain some microbe that can kill the mold, or that would guarantee that you will not get mold in the place you spray it, like an insulating material}. O2 goes far in speculating on the future of bio-technology, discussing how artificial organs and artificial meat will be grown in labs - bringing big positive aspects, and how research on genetic engineering can revolutionize reproduction, where individuals might even be able to self-reproduce. According to her, in this last instance, ethics will be needed to regulate such progress. O3 is also positive and excited about the future of bio-technology, where he believes it will help develop tree species that grow faster, and that can be used in construction, as well as other materials that \textit{we can't even imagine now}.

\section{Discussion}
\label{Discussion}

Using 3D printed cores in the shape of minimal surfaces with mycelium filling is a method to combine the good mechanical properties of various minimal surfaces with the documented good acoustic and thermal insulating properties of mycelium. In this way, 3D-printed minimal surfaces can be used as reinforcement for mycelium-based composites. The mycelium bounded with the wood-based 3D printing filament we used, meaning this is a method to create solid components. the ordered complex geometries of the minimal surfaces can also help mycelium grow faster, as \cite{alima2022bio} noticed in their work.
Working in a design studio setting rather than a laboratory setting, the conditions for conducting the study did not follow precise protocols that are common in biological or structural engineering research. This means we did not measure the amount of mycelium placed on/inside each brick. A future study focused on the material properties of such elements should measure both the material used for the 3D printed core, as well as the mycelium filling to be able to compare such elements. Moreover, participants had the freedom the experiment with the minimal surfaces they wished and did so based on personal and aesthetic preferences. A systematic investigation of each minimal surface type, with different voxel densities and combined with structural testing can be illuminating in understanding the interplay between the 3D printed brick before and after being impregnated and this is what we consider the next step for this research. Mold set in on almost all the bricks after being kept in storage and it was reported by a number of other studies in the field \cite{ciganik2023, appels2019fabrication, Kirdk2022, aiduang2022}.

Most of the participants in the workshop describe designing with mycelium using words such as \textit{interesting, unusual, fantastic, exciting} and even \textit{revolutionary} or an \textit{eye-opening experience}. Many consider the mycelium \textit{friendly}, and easy to work with. The fact that mycelium is a living material means designers need to give away some of the control they are used to having with inert materials. This has been touched upon, for example by \cite{Karana2018, Lim2021, Tamke2012} who talk about changing paradigms where - designers should enter collaborations with their materials rather than try to impose certain shapes on them. Around 20\% of participants were more ambivalent about giving away the control of how the material might behave and look in this material-based design process. More interestingly, some (around 15\%) also describe \textit{being afraid} of the mycelium, or it being \textit{scary} to interact with. One workshop participant reported an itchy skin after interacting with the mycelium, and also being bothered by its smell, meaning mycelium could be an allergen for some. This aspect requires further investigation. Compared to inert materials,  participants appear to have stronger feelings towards a living material such as mycelium. These feelings can be related to biophilia - the well-known theory according to which people have a tendency to seek connections with nature and other forms of life \cite{Wilson1986}. Biophilia is commonly discussed in connection to architecture, and biophlic design is now an established field \cite{Kellert2008}. However, some of the participants also appear to have opposite feelings, they declare feeling scared to interact with or feel afraid of the mycelium. This corresponds to biophobia \cite{SOGA2023, Soga2024}, a newer theory according to which: \textit{biophilia represents just one facet of our relationship with nature and people can also harbor strong negative emotions and attitudes towards nature, increasingly referred to as “biophobia”} \cite{SOGA2023}. Biophobic feelings appear to be on the rise, especially in more urbanised and economically developed societies \cite{Soga2024}. Biophilia is often seen as important in fostering support for environmental conservation \cite{SOGA2023, barragan2022human} while bio-phobia might have the opposite effect, as it can reduce people’s support for pro-biodiversity policies and actions, and increase their antagonism towards nature \cite{Soga2024}. Participants reported that while working with living material, they were required to put in more \textit{care}, and many appeared to enjoy this. Care has been widely discussed as a matter of environmental sustainability, starting with feminists such as d'Eubonne \cite{dEaubonne2022}. Practices of care can contribute to more sustainable behaviors, as a lack of empathy towards, and care for nature is in part responsible for unsustainable behaviors, and climate-change disconnect \cite{Horvath2023a, Olsen2023, Kagan2017, Bentz2020, Bellacasa2017}. All of the participants who describe being afraid of the mycelium also report overcoming the fear after interacting with it. Therefore, experientially working with living materials in design studio settings can be an important way to train a future generation of designers to \textit{come to care through design} and equip them with a more-than-human designerly mindset, a mindset that has been deemed important \cite{COTSAFTIS2023, Wakkary2021, Horvath2023}.

Audience members who took part in the exhibition discussed perceived \textit{challenges}, \textit{ethics} and \textit{bio-technological futures}. 
While all appeared enthusiastic about the future of bio-technology, there are conflicting views about the ways in which this kind of research should be approached. For some, excessive focus on ethics and regulation is detrimental to progress and research, for others ethics should be taken into account only as research ore development can influence human health (meaning all development centers the human in relationship to other species), while finally, for others it is paramount that designers think through ethical questions when they initiate collaborations with living materials, and that there is generally a need to decenter the human in design. However, ethics remains elusive to define, deeply rooted in cultures and \textit{pragmatic} ethics can be difficult to reconcile with \textit{ideal} ethics.

\section{Conclusion}
\label{conclusion}

Mycelium will bind to objects 3D printed using wood-based filaments, and in the case presented in here, compact elements that combine the good mechanical properties of minimal surfaces with the good insulating properties of mycelium were created. However, mold remains a challenge for mycelium-based composites. Material experiences of designers highlight that feelings associated with living materials are stronger than towards inert materials and designers experienced both biophilia, and to a lesser extent biophobia towards mycelium, although all overcame their biophobia in the end. The general public appear positive and excited about the future of bio-technology and its impact on everyday life, but have divergent opinions on how much such research should be regulated by ethical considerations.

\section{Acknowledgements}

This research was partly funded by the Romanian Ministry of Innovation and Digitalization through grant number: PN-IV-P2-2.2-MCT-2023-0037. The exhibition was partly sponsored by: SC Signmaker SRL and SC TAM Wood SRL and the space was granted by Primăria Cluj-Napoca - we are grateful for their support. Furthermore, we thank the interview participants for taking part in our study. Last but not least, we would like to thank the following students who participated in the workshop for their hard work and enthusiasm: Achimeț Clarissa Maria, Barbu Teodora, Benedek Beatrice Mihaela, Bodiu Mara, Brătulete Diana, Brânzan Andrei Lucian, Bungărdean Dan I., Ccuceu Carina, Dobrițoiu Cristina T., Donisa Cosmina Eliza, Floreanu Oana, Gireadă Emilia Gabriela, Halmaghi Ioana, Husariu Delia Monica, Ilea Andreea Florina, Iura Lucian, Iurciuc Ilie, Luca Beatrice Ana-Maria, Macarie Ilinca, Muscar Rareș, Mușină Andreea Bianca, Popa Nicoleta, Pindic Bianca Elena, Secher Leonid, Simofi Istvan, Stoican Denisa, Suciu Georgiana Maria, Șular Florin Vasile, Surdu Teodora Maria, Tăbăcaru Teodora, Vajda Kincso, Vari Denisa Brigitta.

\bibliographystyle{elsarticle-harv} 
\bibliography{Bio-crafting}

\end{document}